\documentclass[conference]{IEEEtran}
\usepackage[utf8]{inputenc} 
\usepackage{commath} 
\usepackage{cite}

\usepackage[pdftex]{graphicx}
\usepackage{array}

\usepackage{url} 
\usepackage{breakurl} 
\usepackage{newunicodechar} 
\usepackage{combelow}
\usepackage{amsmath} 
\usepackage{nameref}
\newunicodechar{ș}{\cb{s}} 
\usepackage{empheq} 

\usepackage{listings} 
\usepackage{booktabs}

\usepackage{alltt}
\usepackage[flushleft]{threeparttable}
\usepackage{multirow}
\usepackage{hhline}

\usepackage{flushend} 

\begin{document}
%
\title{The Potential of the Intel\textsuperscript{\textregistered} Xeon Phi\texttrademark\\ for Supervised Deep Learning}

\author{\IEEEauthorblockN{André Viebke and Sabri Pllana}
\IEEEauthorblockA{Linnaeus University, Department of Computer Science,\\
	351 95 V\"{a}xj\"{o}, Sweden\\
  Email: av22cj@student.lnu.se, sabri.pllana@lnu.se
  }
}

\IEEEspecialpapernotice{\footnotesize(HPCC 2015, \copyright IEEE)}

\maketitle

\begin{abstract}
Supervised learning of Convolutional Neural Networks (CNNs), also known as supervised Deep Learning, is a computationally demanding process. To find the most suitable parameters of a network for a given application, numerous training sessions are required. Therefore, reducing the training time per session is essential to fully utilize CNNs in practice. While numerous research groups have addressed the training of CNNs using GPUs, so far not much attention has been paid to the Intel Xeon Phi coprocessor. In this paper we investigate empirically and theoretically the potential of the Intel Xeon Phi for supervised learning of CNNs. We design and implement a parallelization scheme named \textbf{CHAOS} that exploits both the thread- and SIMD-parallelism of the coprocessor. Our approach is evaluated on the Intel Xeon Phi 7120P using the MNIST dataset of handwritten digits for various thread counts and CNN architectures. Results show a 103.5x speed up when training our large network for 15 epochs using 244 threads, compared to one thread on the coprocessor. Moreover, we develop a performance model and use it to assess our implementation and answer \emph{what-if} questions. 
\end{abstract}

\IEEEpeerreviewmaketitle

\section{Introduction} 
\label{introduction}

\emph{Deep Learning} algorithms are becoming a core component of many modern applications including: self-driving cars\cite{89_self_driven_car}, classification of liver diseases\cite{97_Deep_learning_based_classification_of_focal_liver_lesions}, and speech recognition \cite{105_Android_Speech}. 


A Convolutional Neural Network (CNN) is a deep architecture inspired by the visual cortex of mammals \cite{110_deeplearning_lenet}. CNNs have shown state-of-the art results in fields of computer vision and speech recognition \cite{87_deep_learning_methods_and_applications}. Before utilizing CNNs they need to be trained. Training of CNNs is supervised, using large datasets of labelled data \cite{87_deep_learning_methods_and_applications}. A popular algorithm used for training is the back-propagation algorithm \cite{111_stanfor_bp}.

Training CNNs is computational intense, and often up to several weeks are required to complete a training session if performed sequentially on a CPU. An \emph{epoch} is an iteration within a training session. In order to find the most suitable parameters for a given application, several training sessions are often required \cite{30_Multi_column_Deep_Neural_Networks_for_Image_Classification}. Furthermore, data dependence among training steps in commonly used algorithms makes it non-trivial to exploit the computational capabilities of modern parallel processing devices \cite{75_Hogwild_A_Lock_Free_Approach_to_Parallelizing_Stochastic}.

Compared to other devices (such as GPUs) used for acceleration of computationally intensive tasks, Intel Xeon Phi~\cite{80_overview_of_programming_for_the_intel_xeon_phi} deserves our attention because of programmability~\cite{dokulil13,pllana09} and portability~\cite{kessler12,sandrieser12,benkner11}. 
However, not much research related to the Intel Xeon Phi and Deep Learning has been done so far. A study by Jin et al. \cite{4_accelerating_pattern} target \textit{unsupervised} learning of Restricted Boltzmann Machines and Sparse Auto Encoders. Evaluation performed on an Intel Xeon Phi 5110P resulted with a speed up of $7$ to $10$ times compared to an Intel Xeon E5620. On the contrary, numerous researchers have targeted training of CNNs using GPUs. For instance, Cireșan et al. \cite{16_high_performance_neural_networks_for_visual_object_classification} achieved a $60$x speed up, and Chellapilla et al. \cite{56_High_Performance_Convolutional_Neural_Networks_for} a $4.1$x speed up, compared to a sequential version executed on a CPU, for the MNIST dataset. To our best knowledge the work presented in this paper is the first study of \emph{supervised} learning of CNNs on Intel Xeon Phi.

In this paper we present our parallelization approach for supervised learning of CNNs - Controlled Hogwild with Arbitrary Order of Synchronization (CHAOS) - that is optimized for the Intel Xeon Phi coprocessor. Thread parallelism is used to divide the input over the available threads, allowing threads to process samples concurrently. We apply SIMD parallelism 
in convolutional layers to the computations of partial derivatives and weight gradients. The evaluation of CHAOS was performed on Intel Xeon Phi 7120P coprocessor using the MNIST dataset. We managed to decrease the training time from 31 hours when trained sequentially on the Intel Xeon E5 processor, to 3 hours when trained on the Xeon Phi 7120P coprocessor. Moreover, we developed a performance model to enable assessment of our implementation, and to perform future predictions. We use our performance model to answer \emph{what-if} questions with respect to the number of threads that goes beyond the number of hardware threads supported in the current generation of Xeon Phi. Main contributions of this paper include,

\begin{itemize}
\item development of CHAOS parallelization scheme for the Intel Xeon Phi;

\item experimental evaluation of our implementation using the MNIST dataset;

\item development and validation of the corresponding performance model.
\end{itemize}

The rest of this paper is organized as follows. Section~\ref{background} introduces CNNs and the Intel Xeon Phi. The design and implementation of our approach is discussed in Section~\ref{designimplementation}. Evaluation of our approach is presented in Section~\ref{evaluation}. Section~\ref{relatedwork} discusses the related work. We summarize the paper and highlight future work in Section~\ref{summaryandfuturework}.


\section{Background}
\label{background}

In this section we discuss Convolutional Neural Networks and the architecture of the Intel Xeon Phi.

\subsection{Convolutional Neural Networks}
\label{cnn}

\begin{figure}[ht]
\centering
\includegraphics[width=\columnwidth]{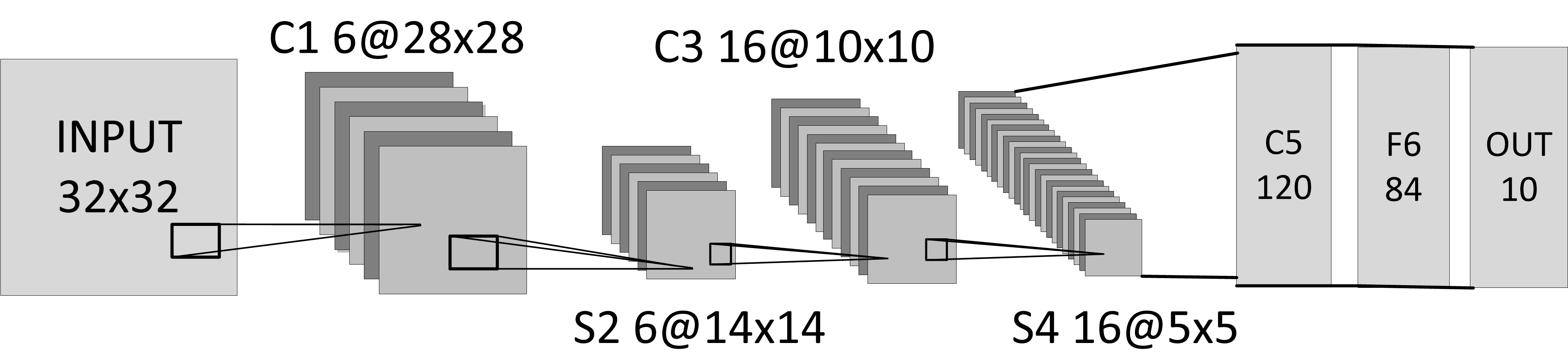}
\caption{The LeNet-5 architecture.}
\label{fig:LeNet}
\end{figure}

A Convolutional Neural Network (CNN) is a variant of a Deep Neural Network introducing two new layer types: \emph{convolutional-} and \emph{pooling-layers}. Inspired by the visual cortex of animals, CNNs are applied to state-of-the-art applications, including computer vision and speech recognition \cite{87_deep_learning_methods_and_applications}. The visual cortex of animals comprise simple and complex cells, which are sensitive to sub-fields of the visual field, called receptive fields. Simple cells are specifically good at detecting edge-like patterns over their receptive fields. Complex cells are locally invariant to the exact position of the detected pattern and span larger receptive fields. These biological properties influenced the creation of CNNs \cite{110_deeplearning_lenet}. Figure~\ref{fig:LeNet} shows LeNet-5 that is an example of a Convolutional Neural Network.

As can be seen in Figure~\ref{fig:LeNet}, each convolutional- and pooling-layer comprise several maps. Neurons in a map cover different sub-fields of neurons in the previous layer - together all neurons in a map span the entire field of neurons in one or many maps in the previous layer. All neurons in a map share the same weight parameters, therefore they extract the same features from the previous layer, although from different parts of the input. Pooling layers aggregate outputs from neurons in the previous layer, thus removing excess information from sequent layers\cite{110_deeplearning_lenet}. 

\begin{figure}[!t]
\centering
\includegraphics[width=0.9\columnwidth]{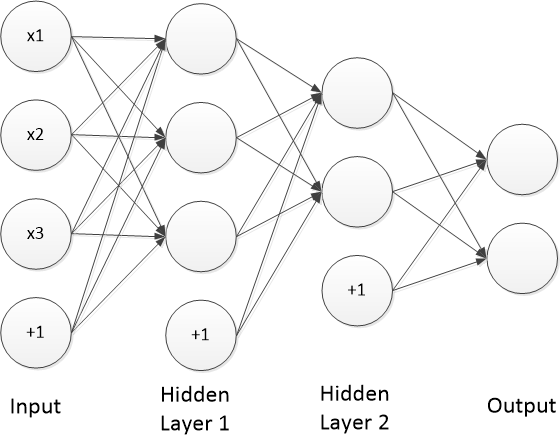}
\caption{A shallow DNN.}
\label{fig:DNN}
\end{figure}

Deep Neural Networks (DNNs) can be visualized as weighted graphs as depicted in Figure~\ref{fig:DNN}. In a nutshell DNNs  are able to make predictions by forward propagating an input through the network. After a forward pass, the output layer contains a vector comprising the prediction. For instance, an image forwarded through the network results in a vector comprising classifications at the output layer \cite{54_Training_an_artificial_neural_network_intro}. 

Starting at the input layer, at each layer the activation for a neuron is calculated as $y^l_i = \sigma(x^l_i)+ I_i^l$, i.e. the output of neuron $i$ at layer $l$ is the input of that neuron sent through an activation function (e.g. sigmoid $\sigma$). $I_i^l$ is the value of the input when no previous layer exists. The input is given by $x^l_i = \sum_j(w^{l-1}_{ji}y^{l-1}_j)$, i.e. the weighted sum of outputs of connected neurons $j$ in the previous layer, multiplied with the weight parameter $w$ connecting neurons $i$ and $j$. This process is repeated at each layer until reaching the output layer. The output vector (values of output neurons) is commonly squashed into normalized values to derive the resulting prediction. In the example of image classification, the output vector contain the probability of the input belonging to a category, e.g. the probability of the image belonging to the category of cats \cite{70_AG_NN,69_stanford_DN}.

Back-propagation is used in \emph{supervised learning} to adjust the weight parameters of the network. The algorithm starts by calculating the error and partial derivatives $\delta^{l}_i$ at the output layer based on the predicted values (from forward propagation) and the true, labelled, value. Then, at each layer, the relative error of each neuron is calculated and weight parameters updated depending on "\emph{how much they were responsible for the error in prediction}". In addition, a decay ($\lambda$) is added to the calculations controlling the impact of the updates. Learning is the process of optimizing the network, making it to adapt to samples seen so far, and thus make better predictions of future, unseen samples \cite{111_stanfor_bp}.

Numerous implementations target Convolutional Neural Networks. \textit{EbLearn} \cite{112_eblearn} from New York University, and \textit{Caffe} \cite{113_caffe} from Berkeley are two examples. In our work we selected a project developed by Dan Cireșan \cite{83_dan_ciresan} with \textit{Boost} library (boost.org) as the only dependency. The implementation target the MNIST dataset of handwritten digits, and has the possibility to define layers, activation functions and connection types dynamically in a configuration file.

\subsection{Intel\textsuperscript{\textregistered} Xeon Phi\texttrademark }
\label{phi}

The Intel Xeon Phi is a many-core shared-memory coprocessor. The one used in our experiments is of type 7120P and comprise 61 cores. The coprocessor hosts a Linux operating system making it possible to communicate with it over \textit{ssh}\cite{66_Optimization_and_Performance_Tuning_for_Intel_Xeon_Phi_part_2}. Two modes are available,
\begin{itemize}
\item \emph{offload}: applications are executed on the host and parts of the code is offloaded on the coprocessor;
\item \emph{native}: code is executed natively on the coprocessor, all code and dependencies have to be uploaded on the device. This mode is used in this paper.
\end{itemize}

\begin{figure}[htb]
\centering
\includegraphics[width=\columnwidth]{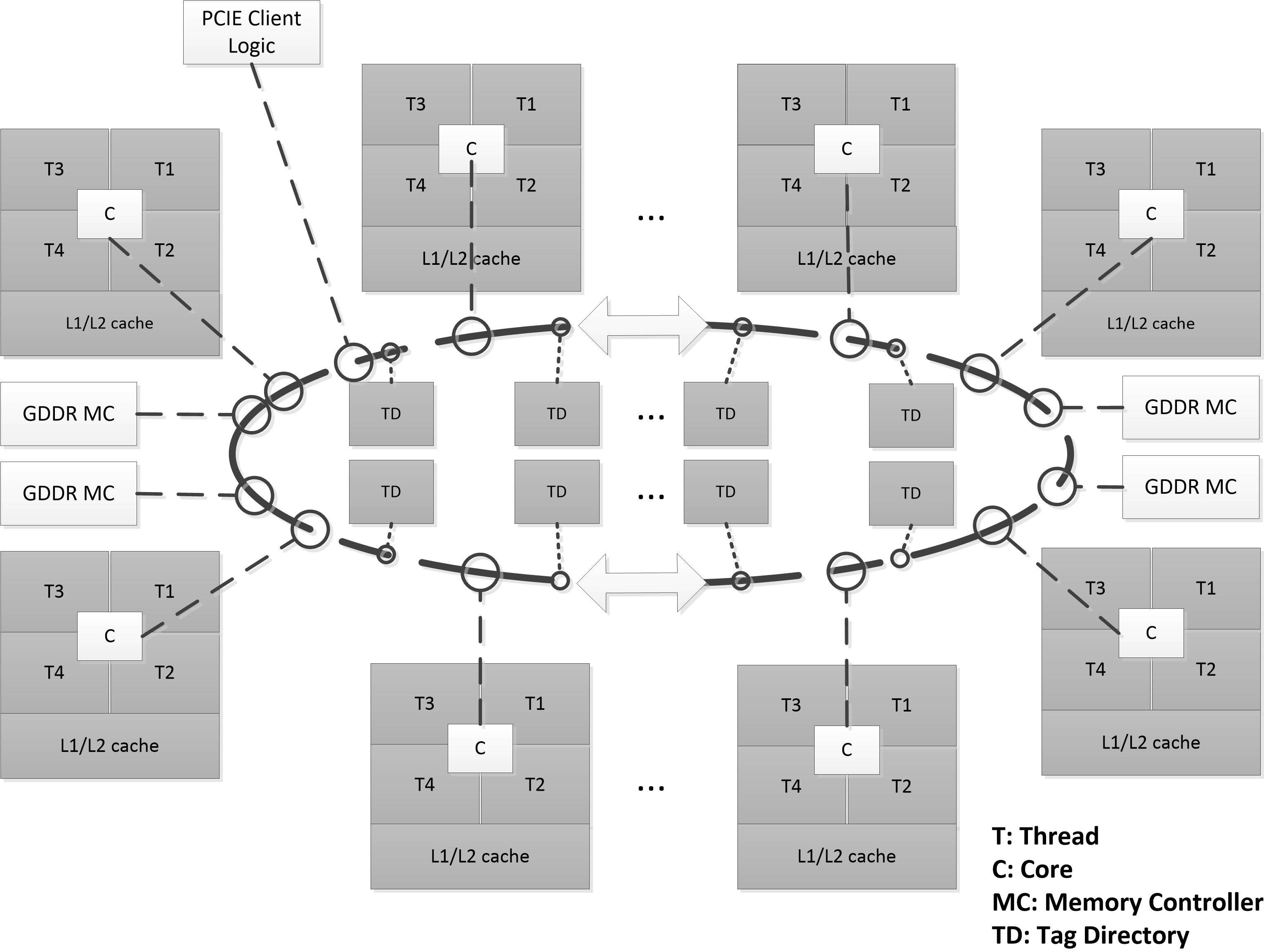}
\caption{The architecture of the coprocessor.}
\label{fig:architecturephi}
\end{figure}

Each core has its own L1 and L2 cache, a clock frequency of 1.2 GHz, and can switch between 4 hardware threads in a round robin manner. In total the coprocessor can manage 244 hardware threads concurrently, thus achieve a double-precision performance of 1.2 teraFLOPS. The L2 cache is 512 KB wide for each core, 30.5 MB in total. The L1 cache comprise 32 KB space for data and instructions respectively for each core. The L2 cache is made coherent through tag directories enabling transfers between core's cache over the interconnect ring. In theory the maximum memory bandwidth is 352 GB/s. The coprocessor used in our evaluation hosts a $\mu$OS of version 2.6.38.8 and a software stack (MPSS) of 3.1.1. An overview of the architecture is shown in Figure~\ref{fig:architecturephi} \cite{66_Optimization_and_Performance_Tuning_for_Intel_Xeon_Phi_part_2,79_intel_xeon_phi_product_brief}.

Data locality and efficient usage of the vector processing units are essential to fully utilize the coprocessor. Through Intel Advanced Vector Extensions (AVX), the 512-bit wide vector processing unit can perform 16 (16 x 32) single-precision, or 8 (8 x 64) double-precision instructions  per cycle \cite{66_Optimization_and_Performance_Tuning_for_Intel_Xeon_Phi_part_2}.



\section{A Parallel Approach for Training Convolutional Neural Networks}
\label{designimplementation}

To explore the potential of the Intel Xeon Phi for supervised Deep Learning we first design and implement a parallelization scheme that we name CHAOS. Thereafter, we perform a platform-independent theoretical analysis of our approach. Finally, we develop a performance model to account for coprocessor-specific hardware characteristics neglected in the theoretical analysis.

\subsection {Parallelization Scheme CHAOS}
\label{sec:chaos}

We designed and implemented a parallelization scheme to exploit the many cores of the Xeon Phi coprocessor titled: \textbf{C}ontrolled \textbf{H}ogwild with \textbf{A}rbitrary \textbf{O}rder of \textbf{S}ynchronization (CHAOS). CHAOS is inspired by existing successful parallelization schemes for stochastic gradient descent. The scheme was adapted to the selected implementation.

\begin{figure}[htb]
\centering
\includegraphics[width=\columnwidth]{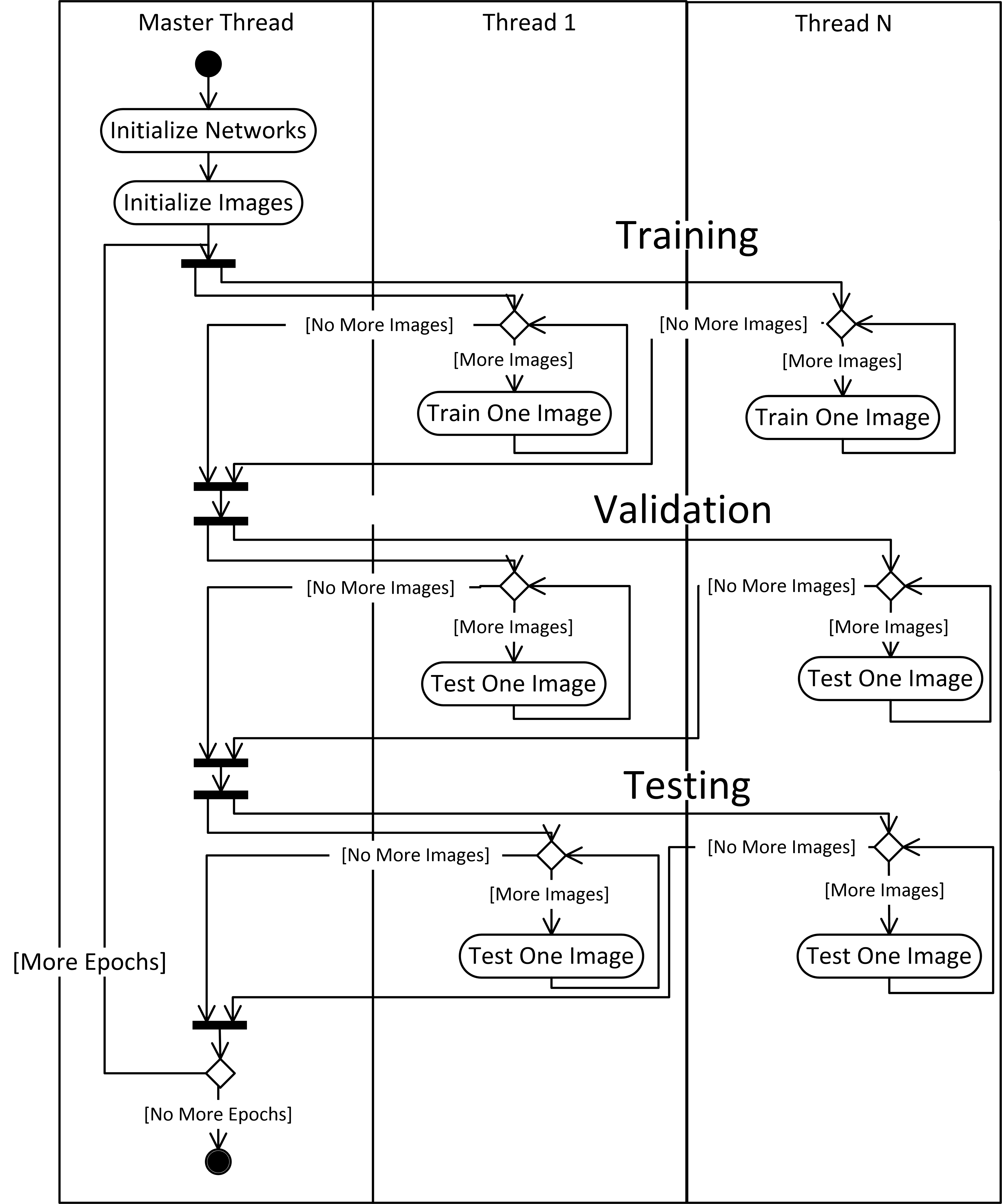}
\caption{An overview of our parallelization scheme CHAOS.}
\label{fig:chaos}
\end{figure}

The key aspects of CHAOS are described in what follows in this sub-section.

\subsubsection{Thread parallelism} Multiple identical network instances (workers) are created, sharing weight parameters; other variables are thread private allowing for several images to be processed concurrently. Figure~\ref{fig:chaos} shows an overview of the scheme, omitting superfluous details. After creating network instances, and preparing images, the training starts. In each epoch three major steps are performed: \textit{Training, Validation}, and \textit{Testing}.

The fist step, \textit{Training}, proceeds with each worker selecting an image in the set of images not yet processed, \textit{forward propagates} it through the network, and calculates the error/loss compared to the labeled value of the image. Thereafter, the partial derivatives are \textit{back-propagated}, adjusting the weight parameters of the network. It is important to note that workers process non-overlapping subsets of images. Because the division of images is non-static, faster workers can process more images than slower workers, reducing the wait time at the end of the work-sharing construct.

After \textit{Training}, each worker participates in the \textit{Validation}-and \textit{Testing}-step to evaluate the network's prediction accuracy. \textit{Validation} operates on the validation set, and \textit{Testing} operates on the test set. In this study we use the same dataset for both training and validation.

Alex Krizhevsky \cite{55_One_weird_trick_for_parallelizing_convolutional_neural_networks} recommends the use of thread parallelism for convolutional layers because they are computational intensive. 

\subsubsection{Controlled Hogwild} Updates of weight parameters in back-propagation are not instant nor significantly delayed. To avoid unnecessary invalidation of cache lines and align memory writes, updates of shared weights are delayed to the end of each layer's computations. Intermediate updates are done to local weight parameters, thus calculating the gradients before sharing them with other workers. The approach is inspired by \textit{HogWild} \cite{75_Hogwild_A_Lock_Free_Approach_to_Parallelizing_Stochastic} proposing instant updates of weights, and delayed updates as proposed by John Langford et al. \cite{76_Slow_Learners_are_Fast}. In our approach, gradients are calculated and saved locally first, however, workers can update the global  weight parameters at any time - they do not have to wait for other workers to finish update before sharing their contributions.

\subsubsection{Arbitrary Order of Synchronization} Because all workers share weight parameters, there is no need for explicit synchronization. However, an implicit synchronization is done in an arbitrary order because writes are performed according to a first-comes-first schedule and reads are performed on demand.

\subsubsection{Vectorization} Additionally, we added SIMD parallelism to the computations in convolutional layers, aligned the memory allocations and memory access to 64 byte. SIMD parallelism in convolutional layers is applied to the computations of partial derivatives and weight gradients, allowing for efficient use of the vector processing unit. An extract from the vectorization report, for the updates of partial derivatives in the convolutional layer, is presented in Listing \ref{lst:bpconvdeltas}. The estimated potential speed up is 3.98.

\begin{lstlisting}[caption={Vectorisation report for the update of partial derivatives in the back-propagation at convolutional layers.},label={lst:bpconvdeltas}]
remark #15475: --- begin vector loop cost summary ---
remark #15476: scalar loop cost: 30 
remark #15477: vector loop cost: 7.500 
remark #15478: estimated potential speedup: 3.980 
remark #15479: lightweight vector operations: 6 
remark #15480: medium-overhead vector operations: 1 
remark #15481: heavy-overhead vector operations: 1 
remark #15488: --- end vector loop cost summary ---
\end{lstlisting}

\subsection {Theoretical Analysis}
\label{sec:tehoreticalanalysis}

The speed up $S_p$ derived in our theoretical analysis is as follows,
\scriptsize
\begin{empheq}[box=\fbox]{flalign*}
S_p = \dfrac{T_1}{T_p} = \dfrac{(a*i + b*it + c) + (d + e*i + f*i + g*it)*ep}{(a*i + b*it + c) + \Big(d+ \dfrac{e*i}{p_i} + \dfrac{f*i}{p_i} +\dfrac{g*it}{p_{it}}\Big)*ep}
\end{empheq}
\normalsize

where $T_1$ is the execution time using one processing unit, $T_p$ is the execution time using $p$ processing units, $a$ and $b$ indicate initializing and preparing images in the memory, $c$ indicates creating network instances, $d$ indicates serialization of intermediate execution results, $e$ is the forward- and back-propagation, $f$ and $g$ indicate forward-propagation,  $i$ is the number of images in the training/validation set, $it$ is the number of images in the test set, $ep$ is the number of epochs, $p$ is the number of processing units, $p_i=min(p,i)$ and $p_{it}=min(p,it)$.

The speed up is defined as the time spent by one processing unit divided by the time spent by $p$ processing units, i.e. how much faster (or slower) $p$ processing units are than one processing unit carrying out the same amount of work. The sequential amount of work carried out by the left term $(a*i + b*it + c)$, preparing the images and network instances, is present both in the denominator and numerator. The sequential work does not have a major impact on the speed up since the constants $a,b$ and $c$ are small in comparison to $e,f$ and $g$. Moreover, $d$ only infer an insignificant amount of work per epoch. Therefore, for an increasing number of images the right term will outgrow the left term in both the numerator and denominator. When increasing the number of processing units the right term in the denominator will decrease and the right term in the numerator remains constant. Since the left term have less impact on the result, the two right terms will control the speed up. Additionally, large epoch ($ep$) counts further increase the right term, and the left term will therefore diminish even further. Nevertheless, the overhead of the sequential work cannot be neglected and will prevent the speed up from achieving a linear behaviour.

The maximum speed up is limited by the number of images in the training set $i$ and test set $it$, both these sets tend to be large, and hence this is a theoretical rather than practical limitation. Additionally, $p_i$ and $p_{it}$ emphasize the difference in semantics of $p$ in their respective context. One processing unit can clearly not process less than one image, therefore we define $p_i \leq i \wedge p_{it} \leq it$. In other words we use $p_i=min(p,i)$ and $p_{it}=min(p,it)$. Furthermore, the speed up is expected to decrease when reaching $p_{it}=it$ as each processing unit tests one image solely at this point. When reaching $p_i=i$ the speed up will not increase further as each processing unit trains and validates one image. 


\subsection {Performance Model}
\label{sec:performancemodel}

The performance model~\cite{pbb08,fahringer04,pllanauml02} enables us to ensure that the implementation behaves as expected. Additionally, it can be used to predict the performance for a varying number of epochs, images and threads. We may use the performance model to answer \emph{what-if} questions with respect to number of threads that goes beyond the number of hardware threads supported in the current generation of Xeon Phi.

\scriptsize
\begin{empheq}[box=\fbox]{multline}
T(i,it,ep,p,s) = T_{comp}(i,it,ep,p,s) + T_{mem}(ep,i,p) \hfill \textbf{(1)} \\ \\ 
= \Bigg(\dfrac{Prep + 4*i +2*it+ 10*ep}{s} \hfill \textbf{(2)}  \\
+ \Bigg(\bigg(\Big(\dfrac{FProp+BProp}{s}\Big)*\dfrac{i}{p_i}*ep\bigg) \hfill \textbf{(3)}
\\ 
+ \bigg(\Big(\dfrac{FProp}{s}\Big)*\dfrac{i}{p_i}*ep\bigg) \hfill \textbf{(4)} \\
 + \bigg(\Big(\dfrac{FProp}{s}\Big)*\dfrac{it}{p_{it}}*ep\bigg)\Bigg) \hfill \textbf{(5)} \\ \nonumber
*CPI\Bigg)*OperationFactor + T_{mem}(ep,i,p)\hfill \textbf{(6)} 
\end{empheq}
\normalsize

where $i$ is the number of images in the training/validation set, $it$ is the number of images in the test set, $ep$ is the number of epochs, $p$ is the number of processing units, $p_i=min(p,i)$, $p_{it}=min(p,it)$, $Prep$ is a placeholder for the sequential amount of work preparing network instances and images, $s$ is the speed of one core on the coprocessor, $FProp$ and $BProp$ are placeholders for the number of operations carried out in forward- and back-propagation, $CPI$ is a factor defining the lowest number of cycles per instruction (in theory) one thread can expect to achieve on the coprocessor, $OperationFactor$ accounts for the approximations done for operations in the calculations. For each CNN architecture size, an approximation of the number of operations is done, replacing the $FProp$ and $BProp$ variables. 

The formula is derived from $T_p$ as defined in section {\ref{sec:tehoreticalanalysis}, before factoring out $ep$ and moving $d$ to the right term. The total execution time is the time consumed by computations $T_{comp}$ plus the time spent waiting for memory operations $T_{mem}$. 

The computational time is calculated as the sum of the sequential work (row 2), plus training, validation, and testing (row 3), (row 4) and (row 5). At row 6 the $CPI$ and $OperationFactor$ adds a penalty to the calculations, and $T_{mem}$ further increase the execution time by adding the memory overhead.

Vectorisation is not explicitly considered in the model, however the $OperationFactor$ will implicitly account for deviations added by vectorisation as well. In essence vectorisation will increase the $CPI$ as fewer instructions are performed per cycle. 

The memory overhead is calculated using the formula,

\begin{equation}
T_{mem}(ep,i,p) = \dfrac{MemoryContention*i*ep}{p}
\end{equation}

The $MemoryContention$ is measured on the coprocessor when several threads fight to access and update the weight parameters in the shared memory space. We performed the measurements for different thread counts and architectures on the Xeon Phi. Each configuration was executed several times and values averaged, deriving the contention expected by $p$ threads processing $p$ images for the given architecture and number of simultaneous threads. Therefore the number of images $i$ is divided by $p$ as we should only expect this contention to occur $i/p$ times.


\section{Evaluation}
\label{evaluation} 

In this section we describe the experimentation environment used for the evaluation and we discuss the obtained results.

\subsection{Experimental Setup}
\label{researchenvironment}

The algorithm is implemented in C++ using OpenMP to exploit thread- and data-parallelism. The application was compiled natively for the coprocessor using the Intel compiler 15.0.0 and the O3 optimization option. All measurements were carried out multiple times, and averaged. As the total execution time is merely a summation of the time spent per epoch, variance in the execution time is mitigated by training several epochs.

The evaluation was performed on an Intel Xeon Phi 7120P coprocessor comprising 61 cores. One core is reserved for software running on the coprocessor, therefore we first restricted our experiments to 60 cores. However, the last core showed an increased speed up in our experiments, and therefore was included in the final results. An Intel Xeon E5-2695v2 with a clock frequency of 2.4 GHz, 132 GB RAM and 48 logical cores was used for comparison of execution times. 

\begin{table}[!t]
\caption{CNN architectures used in evaluation.}
\centering
\begin{tabular*}{\columnwidth}{p{0.6cm}p{0.5cm} r r r r r}
\toprule 
& Type & Maps & Map Size & Neurons & Kernel Size & Weights  \\ 
\midrule 
\multirow{7}{*}{Small} & Input & - & 29x29 & 841 & - & - \\
& Conv & 5 & 26x26 &3,380 & 4x4 & 85 \\
& Max & 5  & 13x13 & 845 & 2x2 & - \\
& Conv & 10 & 9x9 & 810 & 5x5 & 1,260 \\
& Max & 10 & 3x3 & 90 & 3x3 & -\\
& Full & - & 50 & 50 & - & 4,550 \\
& Output & - & 10 & 10 & - & 510 \\
\midrule 
\multirow{7}{*}{Medium} & Input & - & 29x29 & 841 & - & - \\
& Conv & 20 & 26x26 & 13,520 & 4x4 & 340 \\
& Max & 20  & 13x13  & 3,380 & 2x2 & - \\
& Conv & 40 & 9x9  & 3,240 & 5x5 & 20,040 \\
& Max & 40 & 3x3  & 360 & 3x3 & -\\
& Full & - & 150 & 150 & - & 54,150 \\
& Output & - & 10 & 10 & - & 1,510 \\
\midrule 
\multirow{9}{*}{Large} & Input & - & 29x29 & 841 & - & - \\
& Conv & 20 & 26x26 & 13,520 & 4x4 & 340 \\
& Max & 20  & 26x26 & 13,520 & 1x1 & - \\
& Conv & 60 & 22x22 & 29,040  & 5x5 & 30,060 \\
& Max & 60 & 11x11 & 7,260 & 2x2 & -\\
& Conv & 100& 6x6 & 3,600 & 6x6 & 216,100 \\
& Max & 100 & 2x2 & 900 & 3x3 & - \\
& Full & - & 150  & 150 & - & 135,150 \\
& Output & - & 10 & 10 & - & 1,510 \\
\bottomrule
\end{tabular*} 
\label{tab:architectures}
\end{table}

We evaluated our approach using 1, 15, 30, 60, 120, 180, 240, and 244 threads, each thread assigned one network instance. We used the MNIST \cite{107_mnist} dataset of handwritten digits consisting of 60,000 training/validation images, and 10,000 test images. The small and medium CNN architectures were trained for 70 epochs, and the large for 15 epochs. Detailed information of the architectures used in the evaluation can be found in Table~\ref{tab:architectures}. The thread affinity of type \textit{scatter} was used, however we also considered \textit{balanced} which yielded similar results, and \textit{compact} which yielded worse. 

\subsection{Results}
\label{results}

This section present results from the experimental evaluation and results derived using the performance model. The sub-sections are organized as follows:
\begin{enumerate}
\item Speed up of the algorithm, compared to one thread on the Xeon Phi (Figure~\ref{fig:speedupPhi}), and speed up compared to the sequential version executed on Xeon E5 (Figure~\ref{fig:speedupE5}).

\item Execution times for all thread counts and CNN architecture sizes on the Xeon Phi, and the sequential version on Xeon E5 (Figure~\ref{fig:executiontime}).

\item The accuracy of the trained networks expressed in terms of incorrectly predicted images. Table~\ref{tab:err_images} shows the total number of incorrect predictions per configuration and the difference compared to the sequential version.

\item Measured execution times in the experimental evaluation compared to predicted execution times using the performance model, shown in Figure~\ref{fig:prediction}.

\item Future predictions using the performance model when scaling the number of epochs, images and threads (Table~\ref{tab:futurepredictions}).
\end{enumerate}

\emph{Phi Par.} indicates that a parallel version is executed on Intel Xeon Phi, whereas \emph{Xeon E5 Seq.} indicates that a sequential version is executed on Intel Xeon E5. 

\subsubsection{Speed Up}

\begin{figure}[htb]
\centering
\includegraphics[width=\columnwidth]{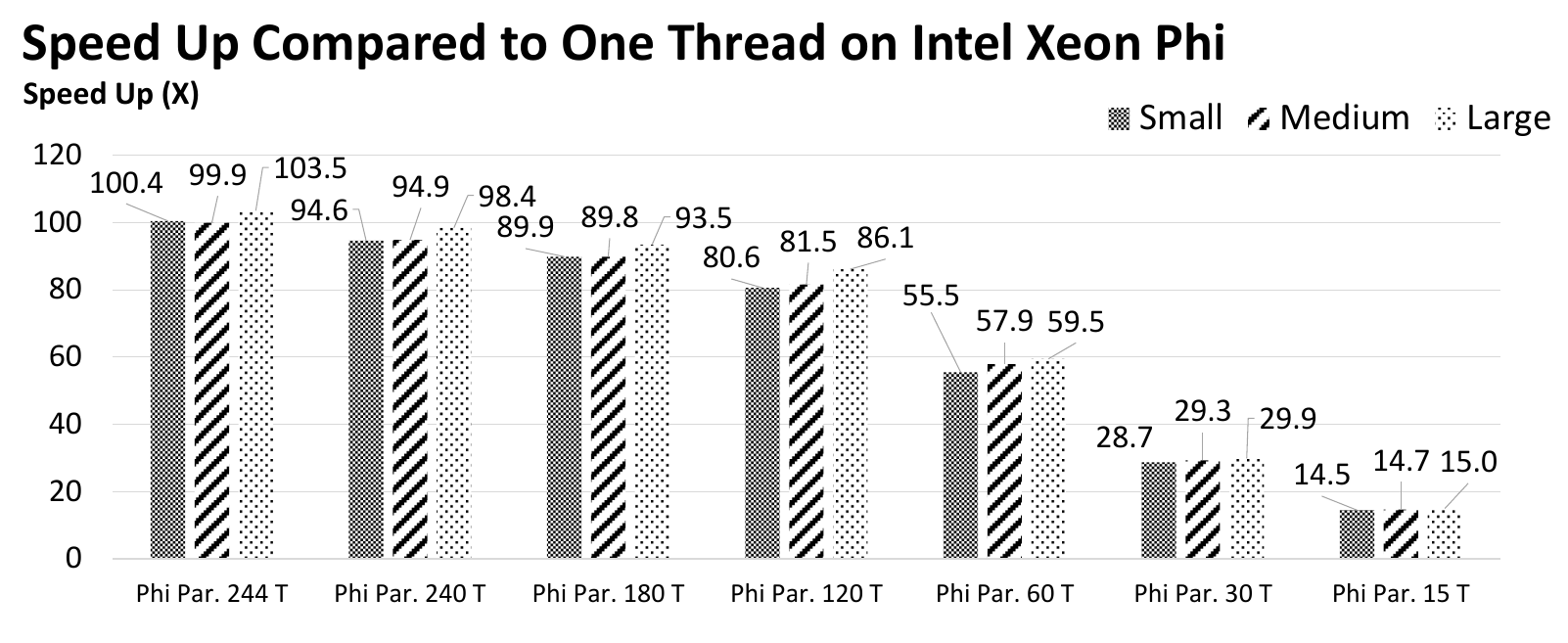}
\caption{Speed up of all CNN architectures used in evaluation for varying thread counts, when compared to one thread on the Xeon Phi.}
\label{fig:speedupPhi}
\end{figure}

Results presented for 244 threads in Figure~\ref{fig:speedupPhi} show that the Xeon Phi yield a $103.5$x, $99.9$x and $100.4$x speed up for the large, medium and small architecture respectively, when compared to one thread on the Xeon Phi coprocessor. Utilizing the last core, otherwise used by the OS, shows a higher speed up than omitting it. Moreover, it can be seen that the speed up is almost linear up to 60 threads were it plateaus - this is true for all CNN architectures. Larger CNN architectures have a slightly higher increase in speed up than smaller ones when increasing the number of threads.

Figure~\ref{fig:speedupE5} shows the speed up compared to the sequential version executed on the Xeon E5. As can be seen, in contrast to Figure~\ref{fig:speedupPhi}, smaller CNN architectures seem to infer a higher speed up than larger architectures for an increased number of threads. Overall, the best speed up is encountered by 244 threads on the Xeon Phi for the large CNN architecture, $14.07$x compared to the sequential version trained on the Xeon E5.

\begin{figure}[htb]
\centering
\includegraphics[width=\columnwidth]{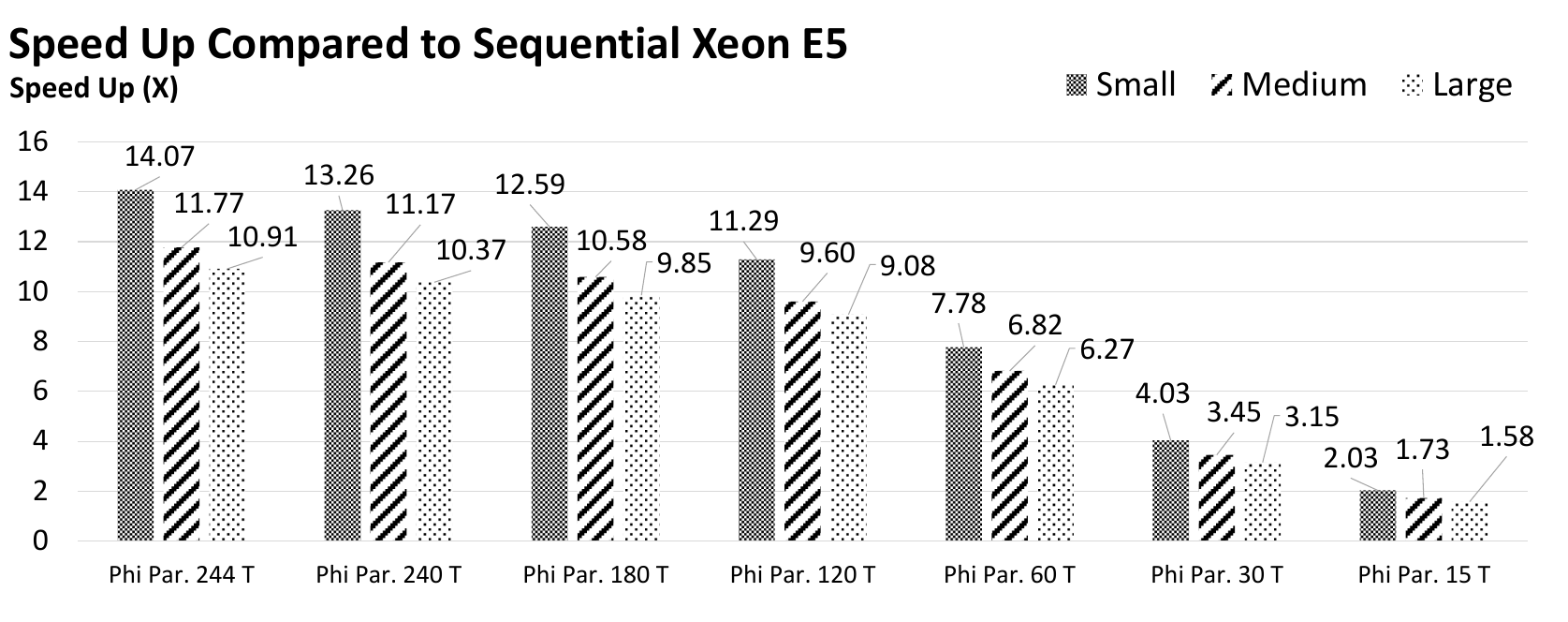}
\caption{Speed up of different CNN architectures and thread counts compared to sequential execution on Xeon E5.}
\label{fig:speedupE5}
\end{figure}

\subsubsection{Execution Time}
The execution times for the parallel version on the Xeon Phi and the sequential version on the Xeon E5 are shown in Figure~\ref{fig:executiontime}. The large CNN architecture, trained for 15 epochs, completes in $31.1$ hours for the Xeon E5 and $2.9$ hours using 244 threads on the Xeon Phi; a \emph{one-and-a-half day} waiting using the Xeon E5 was reduced to \emph{an afternoon} using the Xeon Phi.

\begin{figure}[htb]
\centering
\includegraphics[width=\columnwidth]{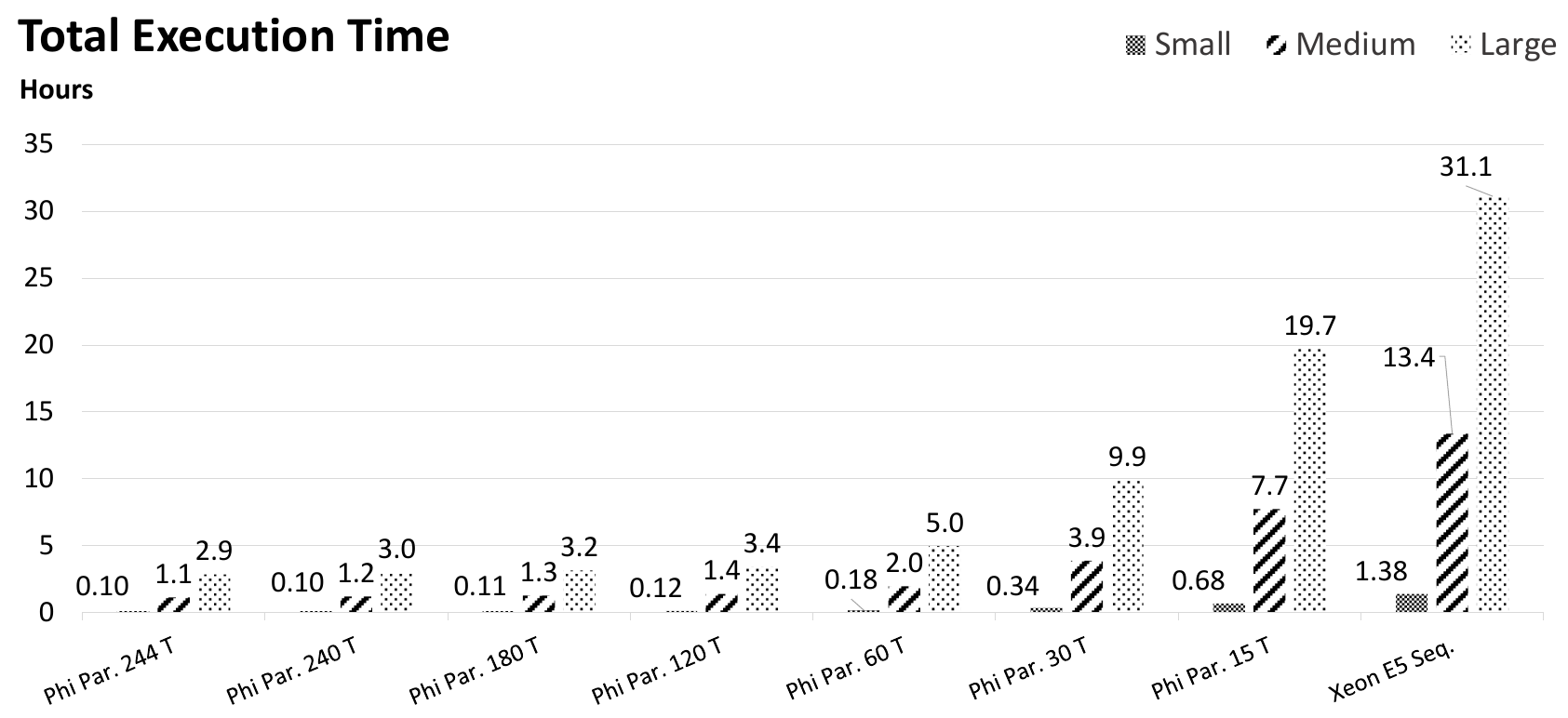}
\caption{The total execution time for the parallel version executed on the Xeon Phi, and the sequential version executed on the Xeon E5.}
\label{fig:executiontime}
\end{figure}

\subsubsection{CNN Prediction Accuracy}
Table~\ref{tab:err_images} shows the number of incorrectly predicted images in total and the difference compared to the sequential version. We may observe that the parallel configurations result in slightly worse predictions in general for the small and medium architecture. For the large architecture, the parallel configurations perform better than the sequential. Nevertheless, the deviation is not abundant in any case and more importantly there is no pattern indicating on a worse prediction accuracy when increasing the number of threads.

\begin{table}[ht]
\renewcommand{\arraystretch}{1.1}
\caption{The number of images incorrectly predicted for the test set in the last epoch. Results are shown for each architecture including the total amount of incorrectly predicted images and the difference to the sequential version on Xeon E5.}
\centering
\begin{tabular}{lrrrrrr}
\toprule
& \multicolumn{2}{c}{Small} & \multicolumn{2}{c}{Medium} & \multicolumn{2}{c}{Large} \\
 & \# Tot & \# Diff & \# Tot & \# Diff & \# Tot & \# Diff \\ \midrule
\textit{Phi Par. 244 T} & 155                        & 2              & 98                          & 3              & 95                         & 1              \\
\textit{Phi Par. 240 T} & 154                        & 1              & 95                          & 0              & 91                         & -3             \\
\textit{Phi Par. 180 T} & 158                        & 5              & 98                          & 3              & 95                         & 1              \\
\textit{Phi Par. 120 T} & 159                        & 6              & 95                          & 0              & 94                         & 0              \\
\textit{Phi Par. 60 T}  & 156                        & 3              & 98                          & 3              & 91                         & -3             \\
\textit{Phi Par. 30 T}  & 156                        & 3              & 98                          & 3              & 90                         & -5             \\
\textit{Phi Par. 15 T } & 153                        & 0              & 100                         & 5              & 84                         & -10            \\
\textit{Xeon E5 Seq.}  & 153                        & 0              & 95                          & 0              & 94                         & 0              \\ \bottomrule
\end{tabular}
\label{tab:err_images}
\end{table}

The overall worst result was encountered by 120 threads on the coprocessor, which incorrectly predicted 159 images out of 10,000 images, 6 images (about 4\%) worse than the sequential version executed on the Xeon E5 that incorrectly predicted 153 images. 

The variance in ending error rates is due to the non-deterministic behaviour of the weight-parameter updates. Moreover, randomization of initial weights and shuffling of images before training was omitted. Nevertheless, we argue that a small deviation in prediction accuracy is an acceptable trade-off for an abundant increase in speed up.

\subsubsection{Measured vs Predicted Execution Time}
Figure~\ref{fig:prediction} present measured execution times (black) and predicted execution times (gray) on the small CNN architecture. The predicted execution times qualitatively match the measured execution times. Lower prediction accuracy can be observed for 15 and 120 threads. The prediction accuracy is better for 180 and 240 threads.

\begin{figure}[htb]
\centering
\includegraphics[width=\columnwidth]{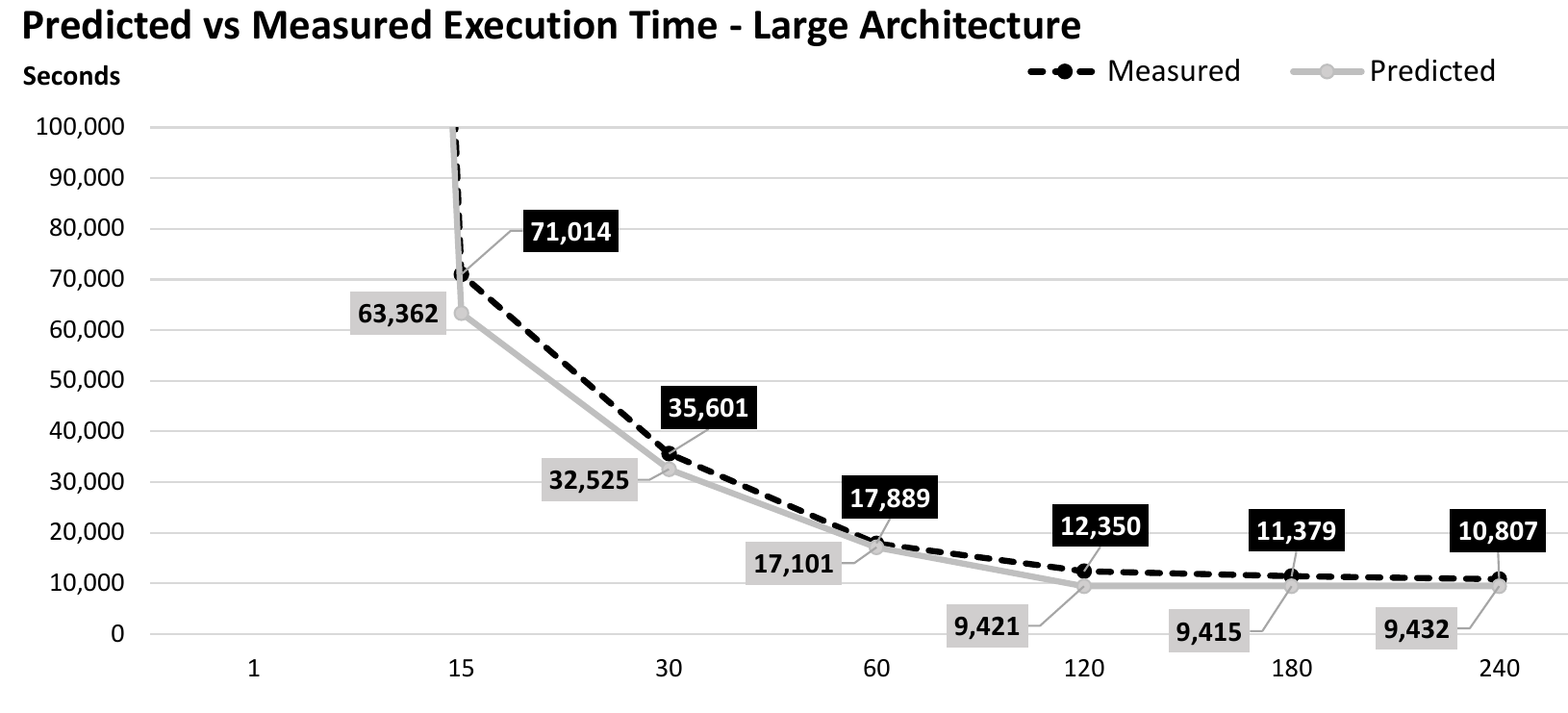}
\caption{Predicted vs measured execution times for the large CNN architecture on the Xeon Phi. Values marked \emph{black} are measured, and \emph{gray} predicted.}
\label{fig:prediction}
\end{figure}

The performance prediction accuracy $\alpha$ is calculated using the following equation,

\begin{equation}
\alpha = \dfrac{\abs{\mu - \psi}}{\psi}100\%
\end{equation}

where $\psi$ is the predicted value and $\mu$ is the measured value. The average accuracy over all thread counts was 15.4\%.

\subsubsection{Performance Prediction}
Table \ref{tab:futurepredictions} shows the predicted execution time (in minutes) if scaling the number of epochs, images, and thread counts for the small CNN architecture. The $i$ is the number of images in the training/validation set, and the $it$ is the number of images in the test set. Doubling the number of epochs or images for a fixed number of threads approximately doubles the execution time. However, for a fixed number of epochs and images, doubling the thread count does not half the execution time.

\begin{table}[ht]
\renewcommand{\arraystretch}{1.1}
\caption{The execution times in minutes when scaling epochs and images for 240 \& 480 threads using the performance model on the small CNN architecture; $i$ is the number of images in the training/validation set, $it$ is the number of images in the test set.}
\centering
\begin{tabular}{r r | r r r r}
\toprule
\multicolumn{6}{c}{240 Threads}\\  
 \midrule
\multicolumn{2}{c}{Images x1000} & \multicolumn{4}{c}{Epochs}\\  
 $i$ & $it$ & 70 & 140 & 280 & 560 \\
 \midrule
 60  & 10  & 8.9  & 17.6 & 35.0  & 69.7 \\
 120 & 20  & 17.6 & 35.0 & 69.7  & 139.3 \\
 240 & 40  & 35.0 & 69.7 & 139.3 & 278.3 \\
\midrule
\multicolumn{6}{c}{480 Threads}\\  
 \midrule
\multicolumn{2}{c}{Images x1000} & \multicolumn{4}{c}{Epochs}\\  
 $i$ & $it$ & 70 & 140 & 280 & 560 \\
 \midrule
 60  & 10  & 6.6  & 12.9 & 25.6  &  51.1 \\
 120 & 20  & 12.9 & 25.6 & 51.1  & 101.9 \\
 240 & 40  & 25.6 & 51.1 & 101.9 & 203.6 \\
\bottomrule
\end{tabular}
\label{tab:futurepredictions}
\end{table}


\section{Related Work}\label{relatedwork}

In this section we highlight representative examples of the work related to (A) machine learning and Intel Xeon Phi, and (B) CNNs and GPUs. To our best knowledge the work presented in this paper is the first study of \emph{supervised} learning of CNNs on Intel Xeon Phi.

\subsection{Machine Learning and Intel Xeon Phi}

Training of Restricted Boltzmann Machines and Sparse Auto Encoders performed by Jin et al \cite{3_training_large_scale_deep_neural_networks_on_the_intel_xeon_phi_many_core_processor} shows $7-10$ times speed up compared to sequential execution on the Xeon E5620. 

A library for Support Vector Machines (SVMs) was developed by You et al. \cite{1_Designing_A_Highly_Efficient_Support_Vector_Machine_For_Advanced_Modern_Multi_Core} to utilize many- and multi-core architectures. The library was evaluated on an Intel Xeon Phi achieving a speed up of $4.4 - 84$ times for various datasets compared to the popular LIBSVM library, when executed sequentially on an Ivy Bridge CPU. 

A Brain-State-In-A-Box neural network was optimized and evaluated on an Intel Xeon Phi 7110P by Khadeer et al. \cite{4_accelerating_pattern} achieving about \textit{two-fold} speed up relative to a CPU with $16$ cores. 

These related works use Intel Xeon Phi for accelerating Restricted Boltzmann Machines and Sparse Auto Encoders, Support Vector Machines, and Brain-State-In-A-Box neural network. In contrast, our work addresses supervised learning of deep CNNs on Intel Xeon Phi.

\subsection{CNNs and GPUs}

Numerous researchers have addressed GPUs and CNNs for the MNIST dataset. Work by Cireșan et al. \cite{16_high_performance_neural_networks_for_visual_object_classification} achieve up to $60x$ speed up when training CNNs on 4 GPUs compared to sequential training on CPU. The GPUs used in the experiments were of type GTX 480 and GTX 580. The CPU was an Intel Core i7-920, 2.66 GHz with 12 GB of memory. 

A state-of-the-art error rate on the MNIST dataset, \textit{0.23\%}, was achieved using a variant of a CNN trained on multiple GPUs\cite{30_Multi_column_Deep_Neural_Networks_for_Image_Classification}. No speed up for a full training session was reported for the MNIST dataset, however, one forward pass of all images in the training set for the Chinese character set required $27$ hours, and hence training one epoch would have lasted for several days. In comparison, on the GPU, training a single epoch of the training set lasted $3.4$ hours. The network was trained for 500 epochs, consequently a CPU was not able to carry out a full training session. 

Chellapilla et al. \cite{56_High_Performance_Convolutional_Neural_Networks_for} investigate the speed up of GPUs training CNNs used for document processing. The MNIST dataset was used for training. Results showed a $4.11x$ speed up for the GPU (Nvidia Geforce 7800 Ultra) compared to the CPU (Intel Pentium 4, 2.5 GHz). 

While these studies target training of CNNs using GPUs on the MNIST dataset, our work addresses training of CNNs on the MNIST dataset using the Intel Xeon Phi coprocessor.


\section{Summary and Future Work} 
\label{summaryandfuturework}

Reducing the execution time for Deep Learning algorithms is essential for their applicability to real-world problems. In this paper we presented a highly parallel approach that exploits thread- and data-parallelism to speed up supervised training of CNNs. Experiments performed on the Intel Xeon Phi 7120P coprocessor demonstrated an increased speed up for larger thread counts. For the large CNN architecture the execution time was lowered from 31 hours on the Xeon E5 to 3 hours utilizing all 244 threads on the Xeon Phi. We used our performance model to evaluate our implementation and to answer \emph{what-if} questions with respect to number of threads that goes beyond the number of hardware threads supported in the current generation of the Intel Xeon Phi. We presume that if more hardware threads become available in the future generations of Xeon Phi coprocessor an even greater speed up could be achieved.

Future work will extend CHAOS to operate in offload mode on multiple Intel Xeon Phis, and evaluate it on the upcoming generation of the Intel Xeon Phi known as \textit{Knights Landing}.


\bibliographystyle{IEEEtran}


\end{document}